\documentclass{ws-rv961x669}
\usepackage{ws-rv-van}     % numbered citation/references (default)
\usepackage{ws-rv-thm}     % comment this line when `amsthm / theorem / ntheorem` package is used
\usepackage{subfigure}     % required only when side-by-side / subfigures are used
\makeindex
\begin{document}

\chapter[Black holes in classical general relativity and beyond]{Black holes in classical general relativity and beyond\label{ra_ch1}}

\author[D. Psaltis]{Dimitrios Psaltis}

\address{School of Physics,\\ Georgia Institute of Technology\\dpsaltis@gatech.edu}

\begin{abstract}
The Kerr-Newman metric is the unique vacuum solution of the General Relativistic field equations, in which any singularities or spacetime pathologies are hidden behind horizons. They are believed to describe the spacetimes of massive astrophysical objects with no surfaces, which we call black holes. This spacetime, which is defined entirely by the mass, spin, and charge of the black hole, gives rise to a variety of phenomena in the motion of particles and photons outside the horizons that have no Newtonian counterparts. Moreover, the Kerr-Newman spacetime remains remarkably resilient to many attempts in modifying the underlying theory of gravity. The monitoring of stellar orbits around supermassive black holes, the detection of gravitational waves from the coalescence of stellar-mass black holes, and the observation of black-hole shadows in images with horizon-scale resolution, all of which have become possible during the last decade, are offering valuable tools in testing quantitatively the predictions of this remarkable solution to Einstein's equations.
\end{abstract}

\body

\section{Introduction}

Black holes are astrophysical objects that have been simultaneously the most unexpected to predict and yet the simplest to describe. Contrary to most other systems in the cosmos, they were first predicted mathematically. Karl Schwarzschild\cite{Schwarzschild1916} derived the vacuum spacetime of a non-rotating object and Reissner\cite{Reissner1916} and Nordstr\"om\cite{Nordstrom1918} extended this work in the presence of electromagnetic fields. What is remarkable about these solutions, albeit not understood until much later, is the fact that they have true physical pathologies, such as curvature singularities. However, these pathologies are surrounded by horizons, which are virtual surfaces from which not even light can escape. As a result, the entire universe outside the horizon is unaware of the presence of these pathologies and immune to them.

Because of this property, the existence of black holes in the real universe was bitterly debated\cite{Thorne1994}. It was only when a pathway to their formation was discovered by Oppenheimer and Snyder\cite{Oppenheimer1939b}, via the collapse of a dust cloud, and other massive compact-object alternatives were shown not to be viable as end states of the collapse of very massive stars\cite{Oppenheimer1939a,Harrison1965} that black holes began to be accepted as plausible objects to search for in the Universe.

Advances in astronomical instrumentation in the 1960's offered tantalizing, albeit circumstantial, evidence that some objects that were newly discovered at the time were indeed black holes. Quasars, the bright sources discovered in the centers of galaxies at cosmological distances\cite{Schmidt1963}, could be understood energetically only if they were powered by accretion onto supermassive black holes\cite{lb1969}. On the opposite end of the mass spectrum, bright X-ray sources discovered in binary systems\cite{Bowyer1965} that were too massive to harbor neutron stars were soon dubbed ``black-hole candidates''\cite{Webster1972}. 

The impetus from these theoretical work and astronomical discoveries led to searches for more vacuum solutions to Einstein's equations, some of which describe compact objects\cite{Stephani2009}. Kerr's solution\cite{Kerr1963} of a rotating vacuum spacetime, as well as its generalization by Newman\cite{Newman1965} in the presence of electromagnetic field, were shown to encompass the earlier solutions by Schwarzschild, Reissner, and Nordstr\"om as limiting cases. Remarkably, among all known solutions, these were the only ones that had their pathologies surrounded by event horizons. The reason for this uniqueness was soon discovered  in a series of articles by Israel, Carter, Hawking, and Robinson\cite{Israel1967,Israel1968,Hawking1972,Robinson1975}. Their work, collectively known as the no-hair  theorems, showed that the only asymptotically flat vacuum solution of Einstein's equations that is free of singularities outside a horizon and free of closed time-like loops is the one described by the Kerr-Newman metric. 

It is important to emphasize here that the no-hair theorem does {\em not} state that the only vacuum solution to the field equations is the Kerr-Newman metric. In fact, there exists an infinite number of them (see, e.g., the so-called Manko-Novikov metrics\cite{Manko1992}, which have been used extensively in recent tests of gravity). The main distinguishing property between the various solutions is that the Kerr-Newman metric has all its pathologies enshrouded by a horizon whereas the others do not. For this chapter, I will refer to objects with no pathologies outside their horizons as ``black holes'' and to all other objects as ``naked singularities''. 

Even within the confines of General Relativity, there are multiple ways to form naked singularities, albeit with ``matter'' fields that do not follow the properties of ordinary matter, that are placed in unphysical configurations, and/or that are finely tuned, making them unstable to perturbations\cite{Choptuik1993,Ori1987,Ori1990,Shapiro1991,Shapiro1992,Gundlach2007,Joshi2011}. Contrary to the case of the Kerr-Newman metric and despite intense analytical and numerical efforts, nobody has found a pathway to forming a naked singularity from stable initial conditions that are astrophysically realistic. If indeed, there is no physical mechanism to form a naked singularity, as posited by Penrose's Cosmic Censorship Conjecture\cite{Penrose1969,Penrose1998}, then all massive compact objects that we observe in the universe must be described by the Kerr-Newman metric. This metric still has substantial pathologies but the presence of horizons surrounding these pathologies protects us from them and enables us to predict the outcome of physical experiments, at least as they take place in our observable universe. Albeit not satisfactory, from a physical point of view, this property allows us to follow the black-hole equivalent of David Mermin's well-known aphorism for quantum mechanics: ``shut up and calculate''\cite{Mermin1989}. In this chapter, I will focus only on the classical description of phenomena that occur outside the event horizons of black holes; the rich phenomenology as well as the multitude of paradoxes one reaches when considering the presence of quantum fields near black-hole horizons can be found in a number of reviews~\cite{Wald2001,Harlow2016,Polchinski2017,Almheiri2021}; see also Chapter~2 of this volume.

The geometry of the Kerr-Newman black hole is fully described by three parameters, which we denote heuristically as the mass, the spin angular momentum, and the charge of the black hole. The case of a charged black hole does not appear to be astrophysically viable, because such an object would quickly neutralize by attracting opposite charges from the surrounding matter\cite{Wald1984} or those created spontaneously because of the presence of strong electric fields outside the horizon\cite{Gibbons1975}. For this reason, I will assume in this chapter that all astrophysical black holes are charge neutral. 

The uniqueness of the Kerr-Newman solution in General Relativity and its dependence on only two astrophysically relevant parameters is the origin of the expected extreme simplicity of black holes. If the theory of General Relativity provides an accurate description of astrophysical black holes and we can prove that these objects have event horizons rather than surfaces, then all gravitational phenomena in their vicinities should be described by the Kerr metric and should only involve lengths and times that simply scale with the black-hole mass (modulo small corrections for the spin). This statement is what makes black-hole tests of gravity very powerful and appealing. If we were to find evidence, through observations and experiments, that the Kerr metric does not provide an accurate description of astrophysical black holes then we would not have the luxury to assign any potential deviations to a putative unresolved gravitational structures (e.g., independent multipole moments, mountains, etc) or to other unknown ``charges''. Instead, we would have to reach one of only three conclusions: that either the theory of General Relativity is not accurate for black holes, or that the supermassive objects we observe have a solid surface, or that they are naked singularities. Either alternative would be the smoking gun for new physics, ready to be discovered.

\section{The Kerr solution in General Relativity}

As with all solutions to physical equations, there is an infinite number of coordinate systems in which one can express them. This issue is of particular relevance to black-hole solutions for which the coordinates that have the simplest, quasi-Newtonian limit and interpretation are usually pathological on the event horizon and vice versa. For this reason, as well as for ease of use in different situations, the Kerr metric has been expressed in a large number of coordinate systems\cite{Visser2007}.

The Boyer-Lindquist coordinates\cite{Boyer1967} $(t,r,\phi,\theta)$ are among the most widely used because, in this representation, the Kerr metric has only one non-diagonal component. When the spin of the black hole is zero, they reduce to the familiar Schwarzschild coordinates and, at very large distances, the three spatial coordinates reduce to the standard spherical-polar coordinates. Nevertheless, these coordinates are pathological on the event horizon of the black hole and are not useful in understanding the properties of matter and electromagnetic fields that cross the horizon.

The Kerr metric is a solution to the Einstein field equation in vacuum
\begin{equation}
R_{\mu\nu}-\frac{1}{2}R g_{\mu\nu}=0\;,
\label{eq:Einstein}
\end{equation} 
where $R$ and $R_{\mu\nu}$ are the Ricci scalar and tensor and $g_{\mu\nu}$ is the metric tensor. Its line element in Boyer-Lindquist coordinates is given by
\begin{equation}
ds^2 = -(1-\frac{2Mr}{\Sigma}) dt^2 - \frac{4 M a r \sin^2 \theta}{\Sigma} dt d\phi+ \frac{A \sin^2 \theta}{\Sigma} d\phi^2 + \frac{\Sigma}{\Delta} dr^2+ \Sigma d\theta^2
\label{eq:KerrBL}
\end{equation}
where $\Delta(r) = r^2 - 2 M r +a^2 $,  $\Sigma(r,\theta) = r^2+a^2 \cos^2 \theta $, and $ A(r,\theta) =\Sigma \Delta +2Mr( r^2+a^2)$. In this equation and throughout this chapter, $G=c=1$, where $G$ is the gravitational constant and $c$ is the speed of light. The Kerr-Newman generalization of the metric requires simply adding the constant $Q^2$ to $\Delta$, where the parameter $Q$ can be identified with the charge of the black hole.

The Kerr metric is not static, as it is not invariant under time reversal, but it is stationary, as it does not depend explicitly on time. It is also axisymmetric. Its metric components depend only on two parameters, $M$ and $a$, traditionally referred to as the black-hole mass and spin (the physical origin of these identifications will be discussed in \S2.2). At the limit $M,a \rightarrow 0$, it reduces to the Minkowski spacetime (albeit in a very unusual representation\cite{Visser2007}), which is asymptotically flat. The metric can be easily generalized to the case of a theory with a cosmological constant, for which the asymptotic behavior is that of the deSitter background\cite{Vaidya1977}.

\subsection{Horizons and Singularities}

As written in the form of equation~(\ref{eq:KerrBL}), the Kerr metric is singular when $\Delta=0$ or $\Sigma=0$. It is straightforward to show that $\Delta=0$ corresponds simply to a coordinate singularity that can be removed by an appropriate coordinate transformation. Indeed, the Kretschmann scalar for the Kerr metric, which is an invariant measure of the curvature of spacetime and is calculated as the contracted product of the Riemann tensor, is\cite{Visser2007} 
\begin{equation}
R_{\alpha\beta\gamma\delta}R^{\alpha\beta\gamma\delta}=\frac{48 M^2 (r^2-a^2\cos^2\theta)}{\Sigma}
\left[\left(r^2+a^2\cos^2\theta)^2 - 16 r^2 a^2\cos^2\theta\right)\right]
\label{eq:Kretschmann}
\end{equation}
and remains finite when $\Delta=0$. On the other hand, when $\Sigma=0$, the curvature of the Kerr metric becomes infinite and, therefore, this corresponds to a true singularity.

The condition for the true singularity, i.e., $\Sigma=0$, is satisfied when $r=0$ {\em and} $\theta=\pi/2$. This condition is impossible to visualize geometrically, if we consider the Boyer-Lindquist coordinates to behave similarly to the traditional spherical-polar coordinates. Exploring the properties of this singularity in a different set of coordinate systems (those in the Kerr-Schild form) leads to the conclusion that, in those coordinates, the true singularity appears to be ring-like\cite{Visser2007}. Note, however, that this is again a coordinate-dependent interpretation of the singularity.

\begin{figure}[t]
     \includegraphics[width=0.48\textwidth]{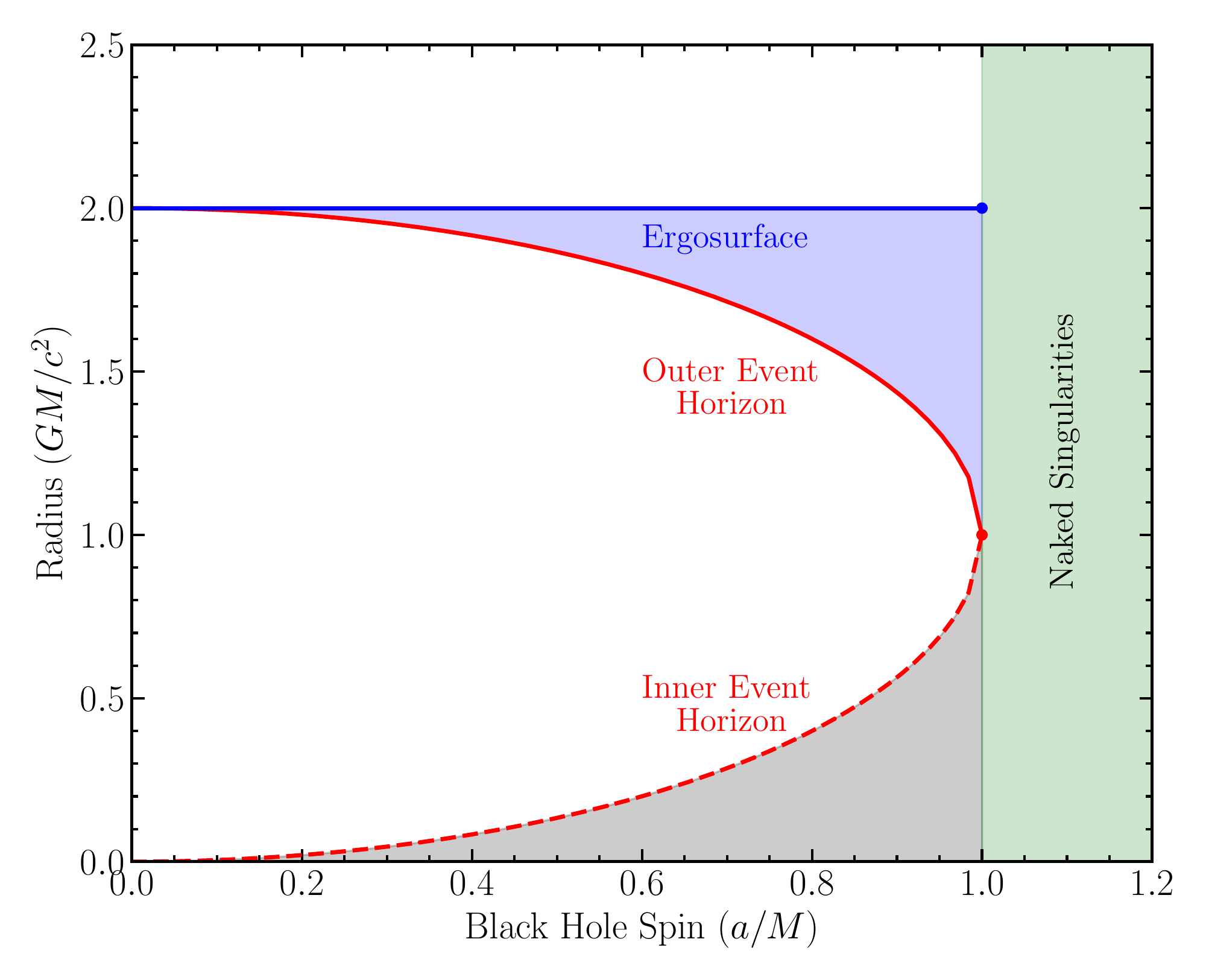}
     \hspace*{5pt}
     \includegraphics[width=0.48\textwidth]{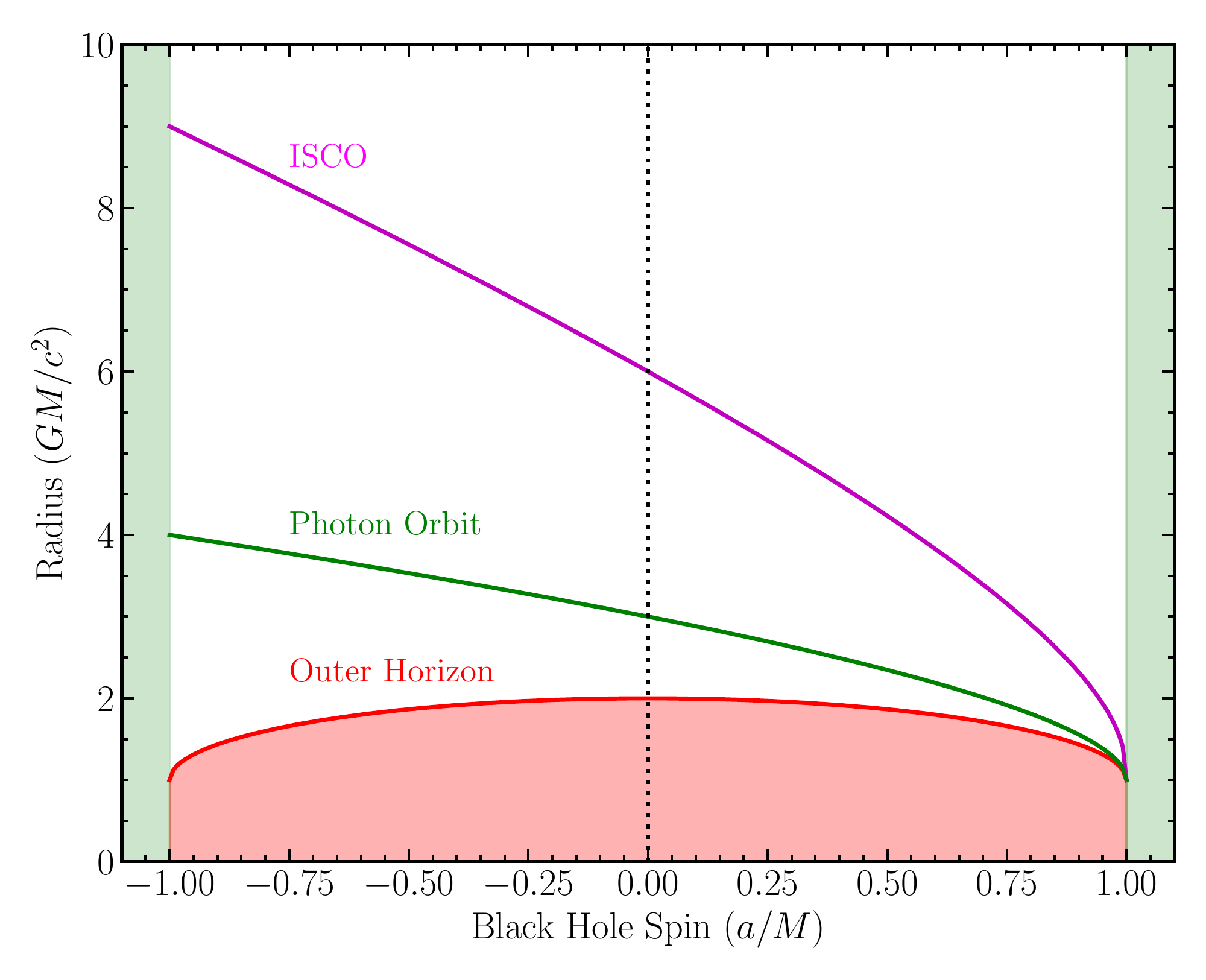}
     \caption{Characteristic radii (in Boyer-Lindquiest coordinates) of the Kerr metric as function of the dimensionless black-hole spin $a/M$. {\em (Left)\/} The radii of the inner and outer horizons as well as of the ergosphere  (see \S\ref{sec:trajectories}). {\em (Right)\/} The radii of the outer horizon, of the photon orbit, and of the innermost stable circular orbit (ISCO). When $a/M>1$, the Kerr metric describes a naked singularity.\label{fig:radii}}
        \end{figure}

When $a<M$, the coordinate singularity at $\Delta=0$ corresponds to two null surfaces at radii
\begin{equation}
r_{\pm}=M\pm \sqrt{M^2-a^2}\;.
\end{equation}
These two surfaces represent the outer and inner horizons of the Kerr metric. The outer (or event) horizon of the Kerr metric is at $r_{+}$ and describes a virtual surface, inside from which not even light can escape to infinity, as the casual definition of an event horizon states. Figure~\ref{fig:radii} shows the dependence of the inner and outer horizon radii on the spin parameter $a$. 

Although the location of the event horizon is at a constant Boyer-Lindquist radius, the geometry of the horizon is {\em not\/} that of a sphere. Indeed, setting $dt=dr=0$, we can write the line element of the 2-surface at $r=r_+$ as
\begin{equation}
ds^2_{r=r_+}= (r_+^2+a^2\cos^2\theta)d\theta^2+\left(\frac{4 M^2 r_+^2}{r_+^2+a^2\cos^2\theta}\right)\sin^2\theta d\phi^2\;,
\label{eq:hor_2sur}
\end{equation}
which is clearly not describing a spherical surface. In fact, the Ricci scalar curvature of this surface is always positive at the horizon but can become negative near the pole when $a>\sqrt{3/4} M$, indicating a rather complex geometry\cite{Visser2007}. The surface area of the horizon is equal to
\begin{equation}
A_{r=r_+}=4\pi\left(r_+^2+a^2\right)\;.
\label{eq:hor_area}
\end{equation}

When $a>M$, the equation $\Delta=0$ has no real solutions and, therefore, the resulting spacetime has no horizon. On the other hand, the singularity at $\Sigma=0$ remains the same and, therefore, the configuration represents a true physical singularity that is not surrounded by a horizon. As discussed in the introduction, although these are valid solution to Einstein's field equations, no scenario is presently known for the formation of such superspinning objects starting from realistic astrophysical initial conditions.

The special case of $a=M$ describes a Kerr black hole with the maximum possible value of the spin parameter $a$. In Boyer-Lindquist coordinates, its horizon is at $r_+=M$. Expanding the expression for the metric to small distances outside the horizon, in the so called Near Horizon Extreme Kerr (NHEK) limit, gives rise to a non-singular metric that has a more extended series of symmetries than the full Kerr solution\cite{Bardeen1999}.

\subsection{Masses and Spins}

The two metric parameters, $M$ and $a$, are typically referred to as the mass and spin angular momentum of the black hole. The reason behind this identification can be seen by taking the asymptotic limit of the metric at large distances ($r\gg M$), i.e., approaching the Newtonian limit. To achieve this, we first set $a=0$, keep only terms of first order in $M/r$, and drop terms of order $(M/r) (dr/dt)^2$, since for massive particles the Newtonian kinetic and potential energies are comparable and, hence, $(dr/dt)^2\sim M/r$. The result is
\begin{equation}
ds^2=-\left(1-\frac{2M}{r}\right)dt^2 +dr^2+r^2 d\theta^2 +r^2\sin^2\theta d\phi^2\;.
\label{eq:kerrBL_assympt}
\end{equation}
Using this metric to write the geodesic equation for the radial motion of a particle, in the same limit, we obtain
\begin{equation}
\left.\frac{d^2 x^\mu}{dt^2}+\Gamma^{\mu}_{\alpha\beta}\frac{dx^\alpha}{dt}\frac{dx^\beta}{dt}=0\right\vert_{\mu=r}\Rightarrow 
\frac{d^2r}{dt^2}=-\frac{M}{r^2}\;,
\end{equation}
which is nothing but the radial equation of motion for a particle in the Newtonian gravitational field of an object with mass $M$. In other words, the motion of a test particle in any gravitational experiment far from the black hole will be identical to that of a test particle in the gravitational field of a Newtonian object with mass $M$. 

Similarly, if we allow for non-zero but small values of the parameter $a$, we can write the metric in the form
\begin{equation}
ds^2=-\left(1-\frac{2M}{r}\right)dt^2 +\left(1+\frac{2M}{r}\right)dr^2+r^2 d\theta^2 +r^2\sin^2\theta \left(d\phi - \frac{2 a M}{r^3} dt\right)^2\;.
\label{eq:kerrBL_a_assympt}
\end{equation}
(Note that this last expression is valid only to first order in $a$).The last term in the line element~(\ref{eq:kerrBL_assympt}) is reminiscent of a coordinate transformation to a frame rotating with an angular velocity $\omega=2 a M/r^3$. We will see later that this is the frequency of rotation of a free-falling observer with zero angular momentum. 

More importantly for our understanding of the metric parameter $a$, we can use equation~(\ref{eq:kerrBL_a_assympt}) to calculate the precession of equatorial, eccentric orbits outside the horizon of a Kerr black hole and find that, during each orbit, their periapses advance by\cite{Boyer1965}
\begin{equation}
\Delta\phi=2 \pi\left(1+3\frac{M^2}{l^2}+4\frac{aM^2}{l^3}+...\right)\;,
\end{equation}
where $l$ is the angular momentum of the orbit. We can now compare this result to the early work by Lense and Thirring\cite{Lense1918}, who calculated the relativistic advance of the periapsis around a uniform sphere of constant density and radius $R$ that is rotating with an angular frequency $\Omega$ and found it to be equal to
\begin{equation}
\Delta\phi_{\rm LT}=2 \pi\left(1+3\frac{M^2}{l^2}+\frac{8 \Omega R^2 M^2}{5 l^3}+...\right)\;.
\end{equation}
Recognizing that $I=(2/5)M R^2$ is the Newtonian moment of inertia of a uniform-density sphere, this allows us to identify the parameter $a$ of the Kerr metric with $a=I\Omega/M$, i.e., the spin angular momentum of the object. Even though the Kerr spacetime is a vacuum solution of the field equations and does not imply that any matter is performing any actual rotation, the fact that the effect described by the parameter $a$ of the Kerr metric is identical to that of the angular momentum content of a spinning object leads us to identify this parameter with the specific angular momentum of the black hole.

It is important to emphasize here that the mass and spin parameters, $M$ and $a$, of the Kerr spacetime describe the external gravitational field of the object and, therefore, are the only ones that can be measured using gravitational experiments. For this reason, the parameter $M$ is referred to as the gravitational (or Kepler) mass of the black hole to distinguish it from the total amount of rest mass that accumulated to create the black hole in the first place, which it is not equal or even uniquely related to. The non equality of the two masses is a property of all relativistic stars and not just of black holes. For example, the gravitational mass of a neutron star, which is the most compact stellar configuration known, is given by
\begin{equation}
M_{\rm G}=\int_0^R 4\pi r^2 \epsilon dr
\end{equation}
whereas the total baryonic mass that resides within the volume of the star is
\begin{equation}
M_{\rm B}=\int_0^R \frac{4\pi r^2 \rho}{\left[1-2m_{\rm G}(r)/r\right]^{1/2}} dr,
\end{equation}
with $R$ being the stellar radius, $\rho$ and $\epsilon$ the rest-mass and energy density at radius $r$, and $m_{\rm G}(r)$ the gravitational mass enclosed inside a sphere of radius $r$. If we were to assume that $\epsilon\simeq \rho$, we would find that the mass the star presents to gravitational experiments, i.e., $M_{\rm G}$, is smaller than the total baryonic rest mass in its volume, i.e., $M_{\rm B}$, by a factor that depends on the compactness of the star, $M/R$, and the particular mass distribution in its interior. 

In the case of a black hole, the relation between the gravitational mass and the total rest mass of the matter that created the black hole depends also on the particular formation path. An example of this can be easily constructed by considering the formation of supermassive black holes via the coalescence of smaller black holes. As described by the {\em area theorem\/} due to Hawking\cite{Hawking1971}, no astrophysically plausible process can lead to a decrease of the horizon area of a black hole. As a result, if two non-spinning black holes (i.e., with $a=0$) with masses $M_1$ and $M_2$ coalesce into a larger black hole, the fact that the horizon area $A_f$ of the final black holes must be larger than the sum of the horizon areas of the two smaller black holes, $A_1$ and $A_2$ implies that the final mass of the black hole must satisfy (see eq.~[\ref{eq:hor_area}])
\begin{equation}
A_f\ge A_1+A_2 \Rightarrow M_f^2 \ge M_1^2 +M_2^2\;.
\end{equation}
If the  two coalescing black holes are of equal mass, $M_1=M_2$, then $M_f\ge \sqrt{2} M_1$. Although $M_f<M_1+M_2$, the gravitational masses of the two coalescing black holes contribute significantly to the gravitational mass of the final remnant in this scenario. In quasi-Newtonian terms, the energy associated with the remaining ``mass'' will be radiated away in the form of gravitational radiation. On the other hand, if $M_2\ll M_1$, then 
\begin{equation}
M_f\ge M_1+\left(\frac{2M_2}{M_1}\right)M_2\;,
\end{equation}
and, therefore, in the limiting case, only a tiny ($2M_2/M_1\ll 1$) fraction of $M_2$ contributes to the increase in the gravitational mass of the final black hole. In other words, the final gravitational mass of a supermassive black hole will be different if it has been formed as a result of the coalescence of a large number of small black holes or as a result of the hierarchical coalescence of pairs of black holes with increasing larger component masses.

\section{Particle and Photon Trajectories in Black Hole spacetimes}
\label{sec:trajectories}

The motion of particles and photons in the Kerr spacetime have a number of characteristics with no Newtonian analogues but with important astrophysical implications.

We will first consider a massive particle (a ``rocket'' or, more generally, an ``observer'') with strong enough propulsion such that it can stay at fixed Boyer-Lindquist coordinates. This implies that the 4-velocity of the particle is $u=(1,0,0,0)$. However, in order for this to be a physically plausible configuration, this velocity vector needs to be timelike, i.e., $g_{\mu\nu}u^\mu u^\nu=g_{tt}<0$, where $g_{\mu\nu}$ are the components of the Kerr metric.  In Boyer-Lindquist coordinates, this can be true only at radii
\begin{equation}
r>r_{\rm E}\equiv M \pm \sqrt{M^2-a^2\cos^2\theta}\;.
\end{equation}
Remarkably, this radius is larger than or equal to the radius of the outer horizon and defines the outer boundary of the so-called ergosphere of the black hole, or ergosurface. By construction, it is physically impossible for a massive particle to remain at fixed Boyer-Lindquist $(r,\phi,\theta)$ coordinates, independent of the magnitudes of forces that act on the particle trying to keep it there. In other words, the future lightcone for the particle does not include the same Boyer-Lindquist $(r,\phi,\theta)$ location but is tilted away from it. However, contrary to the case of the outer horizon, at which the future lightcones are tilted inward, i.e., towards smaller values of the $r$ coordinate, the lightcones in the ergosphere are tilted azimuthally. The presence of particles or magnetic fields in the ergospheres of black holes is intimately related to mechanisms for extraction of angular momentum from the black hole via the Penrose\cite{Penrose1971} and Blandford-Znajek\cite{Blandford1977} processes, respectively (see also Chapter~7 for further discussion).

A second way to explore the same peculiarity of the Kerr spacetime is by calculating the trajectory of a particle at the equator that is falling from infinity with zero angular momentum. Because the Kerr metric, in Boyer-Lindquist coordinates, is independent of $\phi$, there is a Killing vector $\eta=(0,0,1,0)$ along that coordinate and a corresponding conserved quantity that we identify with the angular momentum of the particle, $l\equiv g_{\mu\nu} \eta^\mu u^\nu=g_{\phi\phi} u^\phi+g_{t\phi}u^t$. For this angular momentum to remain equal to zero, given that neither $g_{t\phi}$ nor $u^t$ are equal to zero, requires that $u^\phi$ is also not zero and, therefore, that the $\phi$-coordinate of the particle changes with a coordinate angular velocity of 
\begin{equation}
\omega\equiv \frac{d\phi/d\tau}{dt/d\tau}=\frac{u^{\phi}}{u^t}=-\frac{g_{t\phi}}{g_{\phi\phi}}\ne 0\;.
\end{equation}
The result is that the free-falling frame of this zero angular momentum observer (ZAMO), which is a Lorentz frame by definition, does not fall radially (as one would expect for a Newtonian object) but rather rotates around the black hole. Because $g_{t\phi}$ is proportional to the spin $a$ of the black hole, it appears as if the spinning black hole drags along these inertial frames. Following similar arguments, we can also conclude that a stationary observer can exist inside the ergosphere only with $\omega>0$, i.e., being dragged along by the black-hole spin as well. 

The Kerr metric in Boyer-Lindquist coordinates is also independent of the $t$ coordinate, leading to a second Killing vector $\xi=(1,0,0,0)$, and a conserved quantity that we identify with the energy of the particle $E=g_{\mu\nu} \xi^\mu u^\nu$. Remarkably, the Kerr metric has another symmetry, this time expressed in terms of a Killing tensor\cite{Walker1970}, which corresponds to a third integral of motion, typically referred to as the Carter constant\cite{Carter1968}. This integral of motion has no obvious Newtonian interpretation, although there exist Newtonian counterparts to it\cite{Will2009}. Together with the condition $g_{\mu\nu}u^\mu u^\nu=-1$, which is satisfied by all massive particles, these four conserved quantities allow for a complete characterization of particle and photon orbits without the need to solve the geodesic equations.

Focusing on circular orbits of massive particles on the equatorial plane, we can use the conserved quantities to calculate their orbital frequencies, orbital velocities, and stability properties. Remarkably, the coordinate angular velocity of equatorial orbits has a very simple dependence on the black-hole properties and the Boyer-Lindquist radius that is reminiscent of Kepler's law:~\cite{Bardeen1973}
\begin{equation}
  \Omega_{\phi}=\left(\frac{1}{M}\right)
  \frac{1}{(r/M)^{3/2}\pm a/M }\;,
\end{equation}
where, hereafter, the $\pm$ sign corresponds to prograde and retrograde orbits. The corresponding orbital period, which sets the fastest dynamical timescale in the vicinity of a black hole horizon, becomes
\begin{eqnarray}
  \tau_{\phi}\equiv\frac{2\pi}{\Omega_\phi}
  &=&8.5\left(\frac{M}{10^9 M_\odot}\right)
  \left[\left(\frac{rc^2}{GM}\right)^{3/2}\pm \frac{a}{M}\right]~{\rm hr}
  \nonumber\\
  &=&0.3\left(\frac{M}{10 M_\odot}\right)
  \left[\left(\frac{rc^2}{GM}\right)^{3/2}\pm \frac{a}{M}\right]~{\rm ms}\;.
\end{eqnarray}
This simple calculation shows that gravitational phenomena can occur at millisecond timescales around stellar-mass black holes and at hour timescales around supermassive black holes.

As in the case of Newtonian orbits, the azimuthal velocities of particles increase with decreasing distance from the black hole. However, at a radius~\cite{Bardeen1973}
\begin{equation}
  r_{\rm ph,\pm}=\frac{2GM}{c^2}\left\{1+\cos\left[\frac{2}{3}\arccos\left(
    \mp\vert \frac{a}{M}\vert\right)\right]\right\}\;,
    \label{eq:rph}
\end{equation}
the azimuthal velocities of the orbits become equal to the speed of light. Inside this radius, which is larger than the radius of the outer horizon $r_+$, no particle can exist in a circular orbit because that would require a superluminal velocity. However, not even all orbits with radii larger than this critical radius are stable. Indeed, there is a characteristic radius for the innermost stable circular orbit (ISCO), inside which radial perturbations render the orbits unstable and cause the particles to plunge towards the horizon. The equatorial ISCO radius for the Kerr metric is given by~\cite{Bardeen1973}
\begin{equation}
  r_{\rm ISCO}=3+Z_2\mp\left[\left(1-Z_1\right)\left(3+Z_1+2Z_2\right)\right]^{1/2}\;,
  \label{eq:risco}
\end{equation}
where
\begin{eqnarray}
  Z_1&=&1+\left(1-a^2\right)^{1/3}\left[\left(1+a\right)^{1/3}
    +\left(1-a\right)^{1/3}\right]\\
  Z_2&=&\left(3a^2+Z1^2\right)^{1/2}\;.
\end{eqnarray}
This radius has significant astrophysical implications in a variety of settings. For example, it represents the point of closest approach during the slow inspiral of binary black holes caused by the emission of gravitational radiation, inside of which the two black holes rapidly coalesce to form the final remnant. In the case of geometrically thin accretion flows, the ISCO radius represents the location at which matter transitions from slowly drifting inwards to plunging towards the horizon. In such systems, the ISCO location has been probed through observations of relativistically broadened iron lines\cite{Tanaka1995,Reynolds2003} and of the temperatures of the thermal spectra of the accretion flows themselves\cite{Shafee2006,Mcclintock2014} (see also Chapter~7 for more discussion).

The critical radius $r_{\rm ph}$ (eq.~[\ref{eq:rph}]) represents the distance from the black hole at which photons trace unstable circular equatorial orbits. The Boyer-Lindquist radii of these orbits depend on whether the angular momentum of the photon trajectory is parallel or antiparallel to the spin angular momentum of the black hole. In fact, there exist spherical photon orbits, i.e., trajectories with a constant Boyer-Lindquist radius but which are not confined on a single plane, for any arbitrary orientation of the photon angular momentum with respect to the spin angular momentum of the black hole. The Boyer-Lindquist radii of these orbits fall between the two limits give in equation~(\ref{eq:rph}). The radii of photon orbits set the frequencies of the gravitational waves during the ringdown of the remnants from the coalescence of binary black holes\cite{Ferrari1984}. Similarly, the radii of photon orbits set the characteristic diameters of the shadows cast by black holes on the optically-thin images of their accretion flows\cite{Bardeen1972}. 

Figure~\ref{fig:radii} shows the dependence of the various characteristic radii discussed here on the spin parameter $a$. Note that, although it appears that the outer horizon, the photon orbit, and the ISCO of a maximally spinning black hole ($a/M=1$) overlap, this is only an illusion due to the nature of Boyer-Lindquist coordinates that become singular near the horizon. In fact, in this limit, the proper distance between these radii tends to infinity\cite{Bardeen1973}.

\section{Multipole expansions, no-hair theorems, and the uniqueness of black-hole solutions}

Soon after the discovery of the Schwarzschild solution, Birkhoff proved that this is the unique spherically symmetric solution to the Einstein field equations, independent of the properties of the matter interior to the domain of solution that sources the gravitational field. This theorem applies to spherically symmetric stars, whether or not their mass distribution is contracting, expanding, or pulsating, as well as to non-spinning black holes, where the source of the gravitational field is hidden behind an event horizon.

Relaxing the assumption of spherical symmetry by incorporating only the effects of spin leads to different generalizations of Birkhoff's theorem\cite{Hartle1968}. Expanded to first order in $a/M$, the external spacetime of a stationary spinning object is again unique and is given by the Kerr metric, expanded to the same order. At the second order in $a/M$, the external spacetime is given by the Hartle-Thorne metric, which depends on three parameters, the mass $M$, the specific angular momentum $a$, and the quadrupole moment of the spacetime $q$. The Kerr metric, expanded to second order in $a/M$, is a special case of the Hartle-Thorne metric with the quadrupole moment set to $q=-a^2 M$. 

In principle, one can continue this approach and expand an arbitrary stationary, axisymmetric, asymptotically flat spacetime that is a solution to the Einstein field equations into a bilinear, infinite series of eigenfunctions\cite{Geroch1970,Hansen1974}. When the spacetime is sourced by a matter distribution (as opposed to being a vacuum black-hole solution), these eigenfunctions can be identified with the multiple moments of the matter field. One set of eigenfunctions with coefficients typically denoted by $M_l$, where $l$ is an even integer, are the relativistic equivalent of the Newtonian mass multipoles. The other set of eigenfunctions with coefficients denoted by $S_l$, where $l$ is an odd integer, represent ``currents'' of the matter field. These do not have a Newtonian analogue but arise from the fact that, in General Relativity, the gravitational field is sourced not only by the distribution of matter but also by the energy associated with its motion. 

The Kerr metric is a special case of this general spacetime with the lowest order mass multipole being equal to the mass of the black hole, i.e., $M_0=M$, and the lowest order current multiple being equal to the spin angular momentum of the black hole, i.e., $S_1=a M$. Because the pair of parameters $M$ and $a$ fully describe the Kerr metric, its remaining multipole moments depend only on these two parameters in a way that can be elegantly expressed by the relation
\begin{equation}
M_l + i S_l = M (i a)^l\;.
\label{eq:nohair}
\end{equation}

The general multipolar expansion of a solution to the vacuum Einstein field equations make it clear that there exist an infinite number of such metrics that are possible within the domain of validity of the theory. When viewed as the external spacetimes of different distributions of matter, this generality is, of course, required in order to accommodate the gravitational fields of complex systems with mountains, valleys, etc. However, when viewed as the complete vacuum spacetimes of black holes, these general solutions are always characterized by one or more serious pathologies: there exist curvature singularities (such as the $\Sigma=0$ singularity in the Kerr metric), closed time-like loops (which will make impossible the solution of a particle trajectory as an initial value problem), or non-Lorentzian signatures (e.g., with more than one time-like coordinates)\cite{Carter1968}. 

For the Kerr-Newman spacetime, all these pathologies are inside the (outer) event horizon and, therefore, do not affect the calculation of physical processes that occur in the observable universe. In fact, as discussed in the Introduction, the Kerr-Newman solution is the unique asymptotically flat, stationary, axisymmetric, vacuum spacetime in General Relativity for which curvature singularities, closed time-like loops, or non-Lorentzian signatures are enshrouded by an event horizon\cite{Israel1967,Israel1968,Hawking1972,Robinson1975}. As a consequence, the external spacetime of a black hole (defined here as a vacuum spacetime with a horizon) in General Relativity, depends only on mass and spin (and potentially charge), with no additional free parameters, or ``hair''. This is the qualitative statement of the so-called no-hair theorem and relation~(\ref{eq:nohair}) serves as its quantitative representation.

A question that arises directly from this discussion is whether it is possible to start from realistic astrophysical initial conditions and reach one of the pathological solutions of the vacuum field equations, i.e., one that has no horizons but exposed pathologies. Perhaps the easiest way to achieve this would be through the gravitational collapse of a distribution of matter with non-trivial multipolar structure, i.e., with mountains and valleys. In this case, however, it can be shown that all information regarding the multipolar structure of the initial configuration is radiated away during the collapse in the form of gravitational waves, leaving behind a simple Kerr black hole\cite{Price1972}. 

Relaxing the assumption of realistic astrophysical initial conditions allows for naked singularities to be formed in idealized situations\cite{Choptuik1993,Ori1987,Ori1990,Shapiro1991,Shapiro1992,Gundlach2007,Joshi2011}. However, none of these configurations, either with respect to the initial conditions or to the stability of the end stage of the collapse, appear to be viable candidates for some of the supermassive compact objects in the universe that we identify with black holes. Such studies have led Penrose to postulate that a cosmic censor somehow always finds a way to hide pathologies behind horizons in our Universe\cite{Penrose1969,Penrose1998}. The nature of this ``censor'', if it exists, remains to be discovered.

\section{Black-Hole Solutions in Modified Gravity Theories}

Black holes described by the Kerr metric are remarkable and unique solutions to Einstein's field equations. However, are their propertied tightly connected to the particular nature of Einstein's theory? Would black holes look different if we introduced a modification to the field equations? 

In order to answer these questions, we need to review first the two ingredients of Einstein's theory of gravity\cite{Will2014}: the equivalence principle and the field equations. The equivalence principle encompasses the requirement that all test particles and photons follow geodesics (i.e., straight lines) in the spacetime and that there is no preferred Lorentz frame or location in the Universe. On the other hand, the field equations provide the recipe for us to calculate the properties of a spacetime as it is being curved by a distribution of matter. 

The easiest way to introduce potential violations of the equivalence principle is to break Lorentz invariance either by imposing a preferred direction in time (as, e.g., in the Einstein-aether theory\cite{Jacobson2001}) or a preferred spacelike foliation (as, e.g., in Horava-Lifshitz gravity\cite{Horava2009}). Spherically symmetric\cite{Eling2006,Barausse2011,Zhang2020} and rotating black-hole solutions\cite{Barausse2016,Adam2022} have been found in such Lorentz-violating theories. The spacetimes of these theories have a number of interesting properties with no General Relativistic counterparts, related to the presence of spin-0, spin-1, and spin-2 polarization modes for the gravitons, each with its own propagation speed and corresponding horizon. Moreover, the presence of a preferred direction in time introduces different causal boundaries beyond the traditional horizons. However, in terms of their predictions for the experimentally accessible motion of particles and photons outside the black-hole horizons, these spacetimes introduce only minor, quantitative, corrections to the predictions in the Kerr metric. Exploring, instead, modifications to the Einstein field equations and searching for black-hole solutions has led to a number of unexpected surprises. 

According to the Lovelock theorem\cite{Lovelock1969}, the Einstein field equations are unique in describing a local, second-order, four-dimensional, metric theory of gravity that involves only a rank-2 tensor field, i.e., the metric. Modifying the field equations, therefore, requires relaxing one of these assumptions. 

The simplest modification to the Einstein field equations was explored by Brans and Dicke\cite{Brans1961}, who introduced to the theory a scalar field with a simple kinetic and potential term that is coupled to the metric through an algebraic function that depends on a single parameter. Surprisingly, the Kerr solution of General Relativity remained valid in Brans-Dicke gravity, independent of the value of the coupling parameter, with the scalar field playing no role in the external spacetime\cite{Thorne1971}. Numerical simulations of the collapse of a distribution of matter into a black hole in Brans-Dicke gravity showed that the scalar ``hair'' are radiated away during the collapse, just like the multipole moments of the spacetime in General Relativity\cite{Scheel1995}. Introducing more general functional forms of the Einstein-Hilbert action (the so-called $f(R)$ theories), or more general scalar, vector, or tensor fields that are coupled to the metric with constant coefficients also led to the same conclusion\cite{Psaltis2008}: that the Kerr metric always remains a solution to these field equations. In fact, for Brans-Dicke or $f(R)$ gravity, it was later shown that the Kerr solution is also unique\cite{Sotiriou2012}.

The fact that the Kerr solution satisfies the field equations of a large class of modified gravity theories can be understood with the following argument. Because the Kerr metric is a vacuum solution in General Relativity, it satisfies the so-called vacuum field equations, i.e., it is Ricci-flat: $R_{\mu\nu}=0$ (cf.\ eq.~[\ref{eq:Einstein}]). However, the Minkowski spacetime, being a vacuum solution itself, is also Ricci flat and satisfies the same equation. As a result, any modification to the theory of gravity that reasonably requires for the Minkowski spacetime to be a solution in completely empty space will also be equivalent to the vacuum Einstein field equation and, therefore, have the Kerr metric as a black-hole solution. 

The only way to introduce modifications to the Kerr metric while retaining the Minkowski spacetime as a solution to the field equations in a modified gravity theory is for the equations to include terms that vanish for a Minkowski spacetime but become non-zero in the presence of a vacuum spacetime with a horizon. For this to occur, of course, the coupling between the metric and the additional fields needs to be dynamical and to depend on the properties of the metric itself.

In dynamical Chern-Simons gravity\cite{Jackiw2003}, which introduces a parity violation term to the field equation, rotating black-hole solutions are indeed different than the Kerr metric\cite{Yunes2009,Delsate2018}. The reason is that both the Minkowksi and the Schwarzschild spacetimes have zero parity and, therefore, the Chern-Simons modification to the field equations identically vanish, rendering them identical to the vacuum field equations of General Relativity. However, the Kerr metric has a preferred orientation and sense of rotation. This is described by the set of non-zero current multipole moments of the spacetime, which cause the Chern-Simons modifications to the field equations to become finite. In order for a black-hole solution to satisfy the new field equations, it needs to be modified with respect to the Kerr metric. 

Similarly, in the so-called Einstein-Gauss-Bonnet theories, the modification to the Einstein-Hilbert action, from which the field equations are derived, involve a Gauss-Bonnet term, i.e., ${\cal G}=R^2-4 R_{\mu\nu}R^{\mu\nu}+R_{\alpha\beta\gamma\delta}R^{\alpha\beta\gamma\delta}$. This term is coupled dynamically to the metric via a scalar field, since a non-dynamical coupling would have caused the corresponding terms in the field equations to vanish identically. The Gauss-Bonnet term vanishes for the Minkowski spacetime, which has no curvature whatsoever. However, in the Kerr spacetime, even though both the Ricci scalar $R$ and the Ricci tensor $R_{\mu\nu}$ vanish, the Kretschmann scalar $ R_{\alpha\beta\gamma\delta}R^{\alpha\beta\gamma\delta}$ is non-zero (see eq.~[\ref{eq:Kretschmann}]). As a result, in this theory, the Minkowski spacetime remains valid but the black-hole solution needs to be modified with respect to the Kerr metric\cite{Kanti1996}.

The above examples describe situations in which the Kerr metric does not satisfy the field equations of a modified gravity theory without modifications. In other words, in these theories, all black hole spacetimes need to be different than the Kerr metric. Alternatively, one can construct a modification such that there is no unique black-hole solution but both the Kerr solution of General Relativity as well as a different one can simultaneously satisfy the field equations. Depending on some black-hole property, such as its mass or spin, the modified black-hole solution might be the energetically favorable one. In scalar-tensor gravity theories, this phenomenon has been referred to as spontaneous scalarization\cite{Silva2018,Doneva2018,Antoniou2018,Dima2020}.

Finally, it is important to emphasize here that the above considerations apply only to the equilibrium (stationary) black-hole solutions in modified gravity theory and their relation to the Kerr spacetimes of General Relativity. If we consider, instead, perturbations of black-hole solutions in modified gravity with additional fields, then the resulting additional scalar, vector, or tensor propagating modes will introduce differences in the generation and propagation of gravitational waves that can be observed during the ring-down phase of black-hole coalescence\cite{Barausse2008} (see also Chapter~8 for additional discussion. Alternatively, if the asymptotic behavior of the spacetime is time dependent, as in the case of scalar-tensor gravity with an asymptotic scalar field evolving cosmologically, then the black hole spacetime itself is modified by the presence of additional scalar ``hair''\cite{Jacobson1999}.

\section{Parametric Extensions to the Kerr metric}

As discussed in the previous section, there exist only limited black-hole solutions in modified gravity theories that can be tested against observations. More importantly, many of these modifications involve discontinuous changes from Kerr. Such discontinuities perhaps make these solutions easily discernible from their General Relativistic counterparts because an observation simply needs to pick one between distinct alternatives. However, because these solutions are often not parametrically different from Kerr, they  prohibit placing observational constraints on modifications that become increasingly better as the quality of observations improve. These arguments have generated the impetus for approaches towards quantifying the properties of a black-hole spacetime using a general scheme that is agnostic to the underlying theory.

In Newtonian gravity, the gravitational potential around a macroscopic object, e.g., the planet Earth, can be written in the most general form in terms of a multipole expansion and observations can be used to measure the coefficients of this expansion. Similarly, in General Relativity, one can write the most general form of the external spacetime of a ``black hole'' in terms of its Geroch-Hansen multipoles and use observations to measure its coefficients and place constraints on potential deviations from the Kerr values (e.g., eq.~[\ref{eq:nohair}]). This approach allows for a general description of a spacetime but only if it  is Ricci flat, i.e., within General Relativity. Perhaps more importantly, because of the no-hair theorem, a general Ricci-flat spacetime expressed in terms of multipoles with arbitrary components will be free of pathologies if and only if the various coefficients are those of the Kerr metric. Any deviation will introduce pathologies in regions that will affect the calculation of observables. This is also true for any other parametrization that enforces Ricci flatness\cite{Collins2004,Glampedakis2006,Johannsen2013a}.

Resolving this issues, i.e., writing a general spacetime that is parametrically different from Kerr while hiding any pathologies behind a horizon requires that we relax the assumption of Ricci flatness. This approach has led to a few types of ``designer'' metrics in which the various metric components have been expanded away from the Kerr forms in terms of functions that are of polynomial nature with arbitrary coefficients\cite{Johannsen2011,Vigeland2011,Johannsen2013b,cardoso2014,Rezzolla2014,Konoplya2016,Carson2020}. In the limit of these coefficients becoming negligible, these metrics tend towards the Kerr spacetime. The requirement that these metrics have horizons that surround any pathologies leads to metric components that are rather complex. In some cases, additional requirements are imposed on the metric properties, such as the existence of three integrals of motion for particles and photons, and introduce additional complexities. Nevertheless, these designer metrics represent the only phenomenological, theory agnostic spacetimes that are regular everywhere outside the black-hole horizons and can be used in testing theoretical predictions against observations of black holes.

\begin{figure}[t]
     \includegraphics[height=2.0in,viewport=0 0 240 425,clip=]{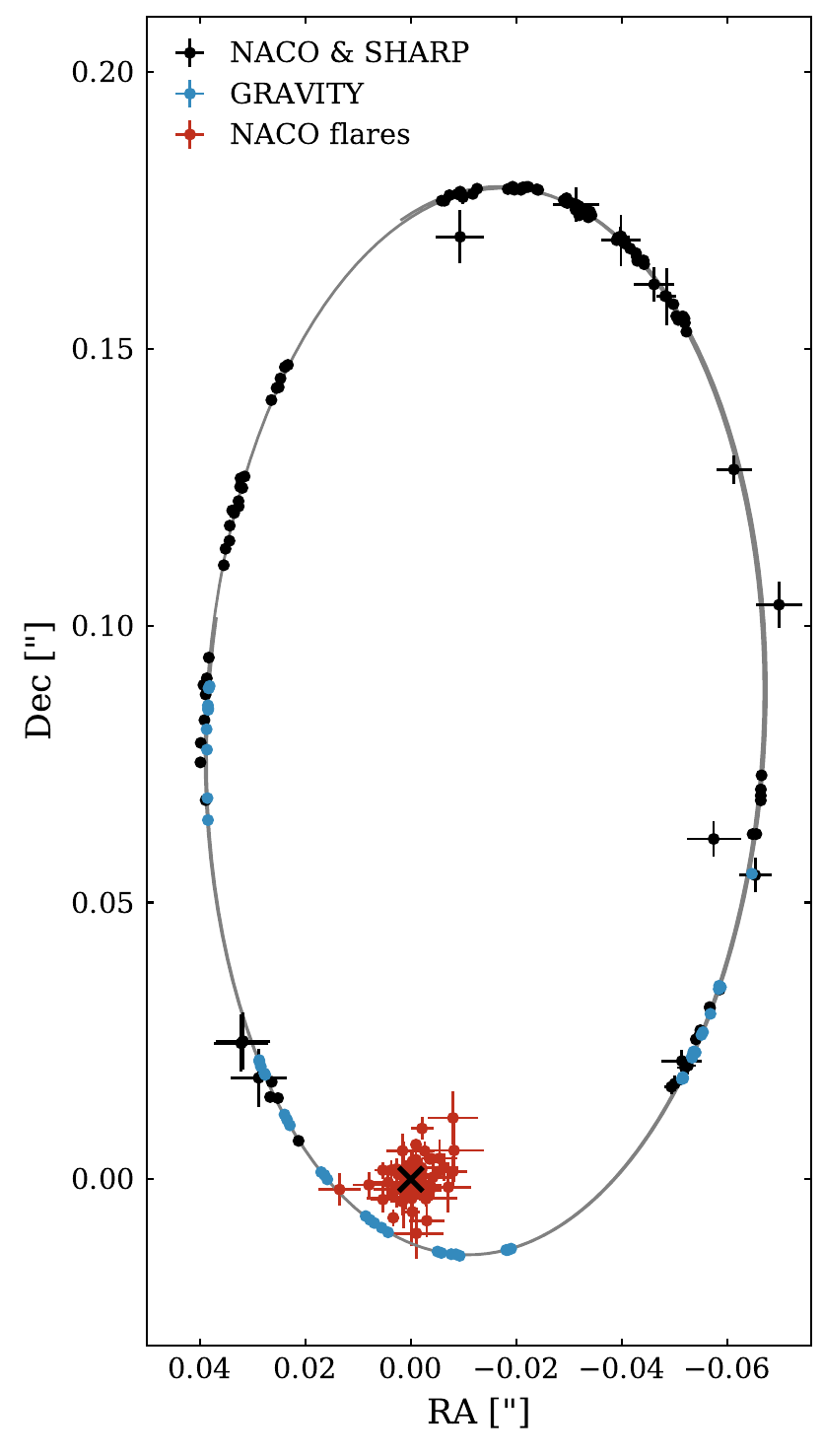}
     \hspace*{5pt}
     \includegraphics[height=2.0in,viewport=0 130 250 390,clip=]{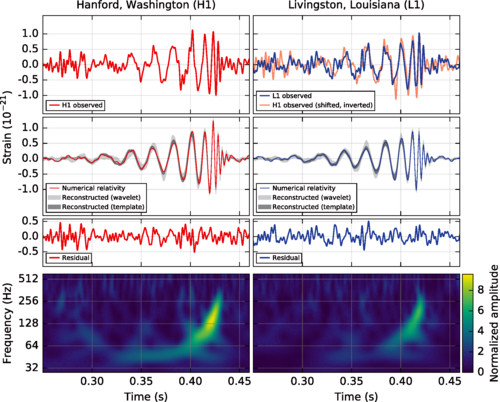}
     \hspace*{5pt}
     \includegraphics[width=1.6in,clip=]{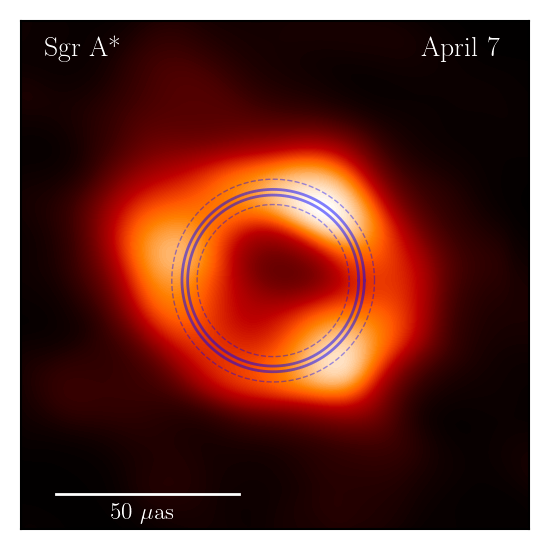}
     \caption{Three observational probes of the Kerr predictions for astrophysical blackholes that have become available in the last decade: {\em (Left)} The periapsis precession of the S0-2 star in orbit around Sgr~A*, the black hole in the center of the Milky Way\cite{Gravity2020}; {\em (Middle)} The gravitational waves observed by the coalescence of two stellar-mass black holes\cite{Abbott2016a}; {\em (Right)} The image of the plasma surrounding the shadow cast by Sgr~A*\cite{EHT2022a}. Each probe is sensitive to a different aspect of the Kerr metric and they have all been shown to be consistent with theoretical predictions at the $\sim 10$\% level.\label{fig:tests}}
        \end{figure}

\section{Testing the Kerr metric}

As mentioned in the Introduction, the uniqueness of the Kerr solution for black holes within General Relativity makes these objects prime candidates for testing the theory itself. Indeed, if astrophysical observations lead to the conclusion that the spacetime of, e.g., one of the supermassive objects in the centers of galaxies, is not described by the Kerr metric then there are only three potential alternatives: {\em (i)\/} the compact object has neither a horizon nor singularities but rather represents a stable configuration of some currently unknown ``matter'' field\cite{Cardoso2019}; {\em (ii)\/} the compact object is a naked singularity; or {\em (iii)\/} General Relativity does not provide an accurate description of black-hole environments and needs to be modified. Whichever alternative turns out to be responsible for such an observation would either require new physics (options (i) in the matter sector and (iii) in the gravity sector) or would allow us to probe close to the singularities that are not surrounded by horizons and, therefore, discover new physics (option (ii)).

Quantitative tests of gravity with black holes have become possible in the last decade, with three different types of observations (see Figure~\ref{fig:tests}). At distances much larger than the horizon radius, the detection of gravitational redshift in the orbit of the S0-2 star around Sgr~A*, the black hole in the center of the Milky Way, led to a very precise measurement of the black-hole mass\cite{Gravity2018,Do2019}. Moreover, observation of Schwarzschild precession in the same orbit between successive periapsis passages resulted in measurements of order unity in the coefficients of the post-Newtonian expansion of the black-hole metric\cite{Gravity2020}. Currently, the magnitudes of these coefficients are consistent with the Kerr predictions, and the upper bounds on any potential deviations are of order unity and limited by systematic effects in the measurements. Successive periapsis passages will only serve to reduce the systematics and tighten the bounds on any deviations. In the future, using a 30-m class telescopes to monitor the orbits of stars that may be detected even closer to the horizon of the black hole has the potential of measuring the spin of the black hole, as well as higher multipole moments of its spacetime\cite{Weinberg2005,Will2008}.

Detecting gravitational waves during the coalescence of binary, stellar-mass black holes with LIGO/VIRGO\cite{Abbott2016b} (see Chapter~8) constrained potential deviations from the predictions of not only the equilibrium metrics of the black holes\cite{Psaltis2021} but also of the generation, polarization, and propagation of the gravitational waves resulting from the collision\cite{Abbott2016b,Abbott2019,Abbott2021}. In the future,  advanced ground-based and space observatories, such as the Einstein Telescope\cite{ET2020}, LISA\cite{LISA2022}, and TianQin\cite{Tian2016}, with allow the measurement of multipole quasinormal modes frequencies during the ringdown of the remnants of black-hole mergers and compare their relative values to the predictions of the Kerr metric\cite{Berti2006,Berti2016}. If LISA detects extreme mass-ratio inspirals (EMRIs), i.e., inspirals of stellar-mass black holes into supermassive ones, then the detection of the frequency evolution of the gravitational waves during the inspiral will lead to a clean mapping of the multipolar structure of the spacetime of the supermassives black holes\cite{Barack2007}. These an other tests of the Kerr metric are described in some detail in Ref.\cite{Arun2022}.

More recently, the observation of black-hole images with horizon-scale resolution from the supermassive black hole in the center of the M87 galaxy\cite{EHT2019a} and Sgr~A*\cite{EHT2022a} using the Event Horizon Telescope (EHT) demonstrated that the inferred sizes of the shadows cast by the two black holes are also consistent with the predictions from the Kerr metric\cite{Psaltis2020,EHT2022b}. The significance of these results lies in the fact that the mass of Sgr~A* and that of the M87 black hole are more than a million and more than a billion times the mass of the Sun, respectively. Together with the constraints imposed by the LIGO/VIRGO observations, they lead us to conclude that there is no experimental evidence to date for any deviation of order larger than $\sim 10-20$\% from the predictions of the Kerr metric across the entire scale of black hole masses currently known in the Universe.

\section{Conclusions}

A little more than a century since their first mathematical description, we have not only understood a lot about black holes and their connection to astrophysics and fundamental physics but have also obtained the first direct confirmations of their quantitative properties through observations. And yet, we have not come any closer to resolving the fundamental problems and paradoxes that arise near and inside their horizons. Addressing them, via advanced observations and a more holistic view of the theory of gravity, is what guides the efforts we are embarking on in the beginning of the second century in our journey of understanding these extreme predictions of Einstein's theory.

\bibliographystyle{ws-rv-van}
\bibliography{chap1}

%\printindex[aindx]                 % to print author index
%\printindex                         % to print subject index
\end{document}